\documentclass[12pt]{article}

\usepackage{graphicx}

\begin{document}

\title{{\small Published in Entropy, {\bf 5}(2003)} \\
Extensive generalization of statistical mechanics based on incomplete information theory}

\author{Qiuping A. Wang \\ Institut Sup\'erieur des Mat\'eriaux du Mans, \\
44, Avenue F.A. Bartholdi, 72000 Le Mans, France}

\date{}

\maketitle

\begin{abstract}
Statistical mechanics is generalized on the basis of an additive information theory for incomplete probability
distributions. The incomplete normalization $\sum_{i=1}^wp_i^q=1$ is used to obtain generalized entropy
$S=-k\sum_{i=1}^wp_i^q\ln p_i$. The concomitant incomplete statistical mechanics is applied to some physical
systems in order to show the effect of the incompleteness of information. It is shown that this extensive
generalized statistics can be useful for the correlated electron systems in weak coupling regime.
\end{abstract}


\section{Introduction}
If the coin of the Chevalier de M\'er\'e\footnote{In 1654, the Chevalier de M\'er\'e, a gamester, proposed to
Blaise Pascal the following problem : two players of equal skill want to leave the table before finishing their
game. Their scores and the number of points which constitute the game being given, it is desired to find in
what proportion they should divide the stakes. In his answer to this question, Pascal laid down the principles
of the theory of probabilities.} comes down heads for 49\%, tails must be 51\% because it has only two sides.
Since Pascal, this hypothesis has never been questioned.

Let us define an ensemble $\Omega$ of N elements (e.g. number of tosses of a coin). Every element has $v$
possible states (sides of coin). A random variable (RV) of this ensemble is denoted by $\xi$ of which the value
is $x_i$ at state $i$ with probability $p_i$. All observed values of $\xi$ constitute an ensemble $\chi=\{x_1,
x_2,...,x_v\}$ with a probability distribution $\emph{P}=\{p_1, p_2,..,p_v\}$. If $v$ is the number of all the
possible values of $\xi$, then $\chi$ is a complete ensemble and $\xi$ is called a {\it complete random
variables} (CRV)\cite{Reny66}. In the toss of a coin, the coin position is a CRV having two values ($v=2$) :
heads or tails. $\xi$ is referred to as independent CRV if all its values are independent (e.g. the result of a
toss has not any influence on the result of the next toss) and exclusive (e.g., heads can not be tails). In
this case, $\emph{P}$ is called a {\it complete probability distribution} (CPD) for which we have the following
postulate :
\begin{equation}                                            \label{1}
\sum_{i=1}^{v}p_i=1.
\end{equation}
The corresponding mathematical framework for calculating $P$ is sometimes called Kolmogorov algebra of
probability distribution \cite{Reny66}. Eq.(\ref{1}) is the foundation of all probabilistic sciences using CPD.
For the average value of certain quantity $\hat{O}$ of the ensemble $\Omega$, we should write
\begin{equation}                                            \label{2}
<O>=\sum_{i=1}^{v}p_iO_i .
\end{equation}
where $O_i$ is the value of $\hat{O}$ in the state $i$.

It is noteworthy that the sum in these equations is over all possible states. Therefore, logically, all
statistical theories constructed within Kolmogorov algebra of CPD should be applied to systems of which all the
possible states are well-known so that we can count them to carry out the calculation of probability or of any
quantity. In physics, this requires not only the exact hamiltonian and solutions of the equation of motion, but
also the mathematical tools allowing to treat exactly the known states in phase space.

However, there are few real systems about which we are capable of treating all information. Many systems in
physics theory are nothing but some isolated and simplified subsystems of the complex and messy world only
partially known. {\it The information needed to specify exactly these subsystems is never completely accessible
for us}. Thus, strictly speaking, the CPD from which a completely accessible information can be calculated is
no more valid. We should address a more general distribution, that is the {\it incomplete probability
distributions} (IPD) due to the {\it incomplete random variables} of which we do not know all the possible
values (i.e. $v$ may be greater or smaller than the real number of possible values). CPD is only a special case
of IPD\cite{Reny66}. Very recently, a possible statistical theory for IPD was
proposed\cite{Wang00,Wang01,Wang02,Wang02a}. This statistics based on the incomplete information consideration
has been applied to a generalization of Boltzmann-Gibbs statistical mechanics (BGS), the so called nonextensive
statistical mechanics (NSM)\cite{Tsal99}. It is shown\cite{Wang01} that this {\it nonextensive incomplete
statistics} (NIS) can indeed overcome some important theoretical difficulties encountered in the last CPD
version of NSM. In addition, the incomplete nonextensive generalization of Fermi-Dirac distribution is proved
to be the only generalized quantum distribution (among several ones) showing the same characteristics as the
distribution of strongly correlated heavy fermions and so is possible to be applied to this kind of
systems\cite{Wang02,Wang02a}.

In this paper, on the other hand, the idea of IPD will be used to generalize BGS for the cases where {\it
additive information and physical quantities} hold. We will show that it is possible to use this extensive
generalization of BGS to describe correlated electrons in the weak-coupling regime.

\section{About incomplete distributions}

A ``ideal coin" has only two sides. So we surely have $v=2$. But what about a coin having thick edge which
possibly remains standing without being observed? And what is the value of $v$ for a quantum coin? For a coin
having overlapped heads and tails? For a coin in fractal or chaotic phase space? These questions may appear
comic. But many real systems can surely be regarded as one of these coins, especially the complex systems with
correlation or effects of correlation that can not be exactly described with analytic methods. For this kind of
systems, the equation of motion becomes incomplete in the sense that some interactions are missing in the
hamiltonian and that its solution can not yield complete knowledge about the system. This incompleteness of
knowledge makes it impossible to apply Kolmogorov theory because Eq.(\ref{1}) and Eq.(\ref{2}) can not be
calculated. In this case, statistical theories for IPD are necessary.

The basic assumption of incomplete statisics\cite{Wang00} is that our knowledge about physical systems and the
relevant probability distributions are incomplete, i.e. $\sum_{i=1}^{w}p_i=Q\neq 1$\cite{Reny66} where $w$ is
only the number of accessible states in phase space. So one can only write $\sum_{i=1}^{w}F(p_i)=1$ where $F$
is certain function of $p_i$. In the case of complete or approximately complete distribution (such as in BGS),
$F$ is identity function. In my previous work\cite{Wang00}, in order to keep the framework of NSM using Tsallis
entropy, I proposed $F(p_i)=p_i^q$ so that
\begin{equation}                                \label{3}
\sum_{i=1}^{w}p_i^q=1,
\end{equation}
where $q$ is Tsallis entropy index\cite{Wang00,Wang01}. Since $p_i<1$, we have to set $q\in[0,\infty]$. $q=0$
should be avoided because it leads to $p_i=0$ for all states. $p_i^q$ can be called {\it effective probability}
which allows to relate the parameter $q$ to observed results and so to interactions. $p_i$ is the {\it `true'
probability} which is physically useful only when $q=1$ for the cases where information is {\it \`a priori}
complete. According to Eq.(\ref{3}), Eq.(\ref{2}) should be written as
\begin{equation}                                \label{4}
<O>=\sum_{i=1}^{w}p_i^qO_i.
\end{equation}

This {\it incomplete normalization} is possible whenever the phase space is partially known or accessible. It
is well known that the systems with complex interactions show often fractal or chaotic phase space, in which a
complete calculation of probability becomes in general impossible because the space is not integrable or
differentiable due to, among others, singular points. In this sense, a possible justification of Eq.(\ref{3})
can be seen in a work of Tsallis\cite{Tsal95} discussing nonadditive energy and probability distributions on
fractal supports, although at that stage the work was not connected to anomalous normalization like
Eq.(\ref{3}). Considering some simple self-similar fractal structures (e.g. Cantor set), one can obtain :
\begin{equation}                                \label{3a}
\sum_{i=1}^{W}[\frac{V_i(k)}{V(0)}]^{d_f/d}=1
\end{equation}
where $V_i(k)$ may be seen as the segments of the fractal structure at a given iteration of order $k$, $V(0)$ a
characteristic volume of the fractal structure embedded in a $d$-dimension Euclidean space, $d_f=\frac{\ln
n}{\ln m}$ is the fractal dimension, $n$ the number of segments replacing a segment of the precedent iteration,
$m$ the scale factor of the iterations and $W=n^k$ the total number of segments at the $k^{th}$ iteration. If
we suppose that the fractal structure with $k\rightarrow \infty$ is a $d_f$-dimension phase space containing
homogeneously distributed points, the {\it complete microcanonical probability distribution} at the $k^{th}$
iteration should be defined as $p_i=\frac{V_i(k)}{V}=\frac{V_i(k)}{\sum_i^WV_i(k)}$ where $V$ is the total
volume occupied by the state points. This distribution obviously sums to one. The problem is that $V$ is an
indefinite quantity as $k\rightarrow \infty$ and, strictly speaking, can not be used to define exact
probability definition. In addition, $V$ is not differentiable (or integrable) and contains inaccessible
singular points. Thus the probability defined above can not be exactly computed. Now if we define
$p_i=\frac{V_i(k)}{V_0}$ as a physical or effective distribution, then we have
$\sum_{i=1}^{W}p_i^{d_f/d}[V_0/V(0)]^{d_f/d}=1$, where $V_0$ is {\it a completely accessible and infinitely
differentiable support} on which the calculation of $p_i$ is possible. If we choose $V_0=V(0)$, i.e. a
$d$-dimension volume containing the $d_f$-dimension fractal volume, we get Eq.(\ref{3}) with $q=d_f/d$. The
conventional normalization $\sum_{i=1}^{W}p_i=1$ can be recovered when $d_f=d$.

The above example is only a case of equiprobable distribution on simple fractal structure, but it illustrates
very well the possibility that, in complex cases, the $physical$ probability may not sum to one and may sum to
unity only through a kind of power normalization which is in addition consistent with the discussions of
reference \cite{Tsal95} on the mass calculation and the information consideration for porous structures.

As mentioned above, Eq.(\ref{3}) and (\ref{4}) have been applied to NSM established with Tsallis entropy
$S_q=-k(\sum_{i=1}^{w}p_i-\sum_{i=1}^{w}p_i^q)/(1-q)$. As a matter of fact, Tsallis entropy implies an
information measure in the form $\frac{(1/p_i)^{q-1}-1}{q-1}$, a nonadditive generalization of Hartley formula
$\ln(1/p_i)$\cite{Wang00}. In this case, the normalization of $p_i^q$ is necessary if we want the entropy to
have same nonextensive properties and variance as generalized Hartley information. Of course, different form of
the function $F$ will lead to different statistics. For example, if we use Hartley formula as information
measure and define expectation value for additive entropy and energy with $F(p_i)$ satisfying $\frac{\partial
\ln F(x)}{\partial x}=\frac{1/x}{\alpha-\ln x-(-\ln Zx)^{1/\gamma}}$ where $\gamma$ is an empirical parameter,
maximum entropy will lead to the famous {\it stretched exponential distribution}\cite{Jund01}
$p_i=\frac{1}{Z}e^{-(\beta E_i)^\gamma}$ where $E_i$ is positive energy of the system at state $i$, $\alpha$
and $\beta$ the Lagrange multipliers related respectively to normalization and energy constraint
$U=\sum_{i=1}^{w}F(p_i)E_i$. When $\gamma=1$, we can recover $\frac{\partial \ln F(x)}{\partial x}=\frac{1}{x}$
as in BGS.

In general, a system becomes nonextensive if it contains interacting parts. But in what follows, we will apply
incomplete normalization to extensive systems. This attempt is partially inspired by the insufficiency of the
nonextensive quantum distributions of NSM to describe correlated electrons in weakly coupling regime, as
discussed in reference\cite{Wang02a}.
\begin{figure}[h] \label{f1}
\includegraphics[width=5in,height=4in]{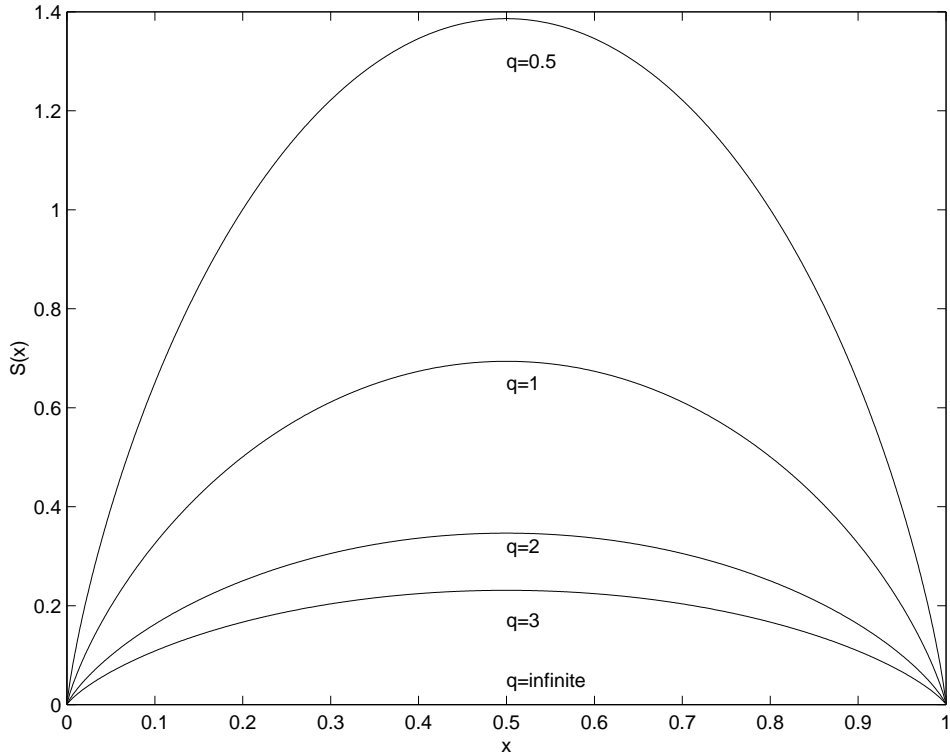}
\caption{$q$-dependence of the concavity of the extensive generalized entropy $S$}
\end{figure}

\section{Extensive incomplete statistics (EIS)}
Now we suppose that the system of interest has $N$ interacting elements and the information $I(N)$ needed to
specify all the elements, as well as the physical quantities, are additive (e.g. $H=\sum_jH_j$ where $H$ is the
hamiltonian of the compound system and $H_j$ that of the $j^{th}$ element). Under these harsh conditions, we
can postulate\cite{Reny66} :

\begin{enumerate}

\item $I(1)=0$ (no information needed if there is only one element)

\item $I(e)=1$ (information unity)

\item $I(N)<I(N+1)$ (more information needed with more elements)

\item $I(\prod_{i=1}^w N_i)=\sum_{i=1}^w I(N_i)$ (additivity of the information needed to specify
simultaneously $w$ subsystems containing respectively $N_i$ elements)

\item $I(N)=I_w+\sum_{i=1}^w p_i^qI(Ni)$ (additivity of information measure in two steps, where
$p_i=\frac{N_i}{N}$ is the probability to find an element in the $i^{th}$ subsystem and $I_w$ the information
needed to determine in which subsystem the element will be found. )

\end{enumerate}

Only the postulate 5 is different from the conventional form because $p_i$ is replaced by $p_i^q$ due to IPD.

As usual, the postulates 1-4 yield Hartley formula $I(N)=lnN$. The postulate 5 becomes $lnN=I_w+\sum_{i=1}^w
p_i^qlnN_i$ which can be recast as

\begin{eqnarray}\label{11}
I_w=\sum_{i=1}^w p_i^qln(1/p_i).
\end{eqnarray}

Now we define an entropy as follows :
\begin{equation}
S=k\sum_{i=1}^w p_i^q\ln(1/p_i)            \label{12}
\end{equation}
which obviously becomes Gibbs-Shannon entropy ($S_{GS}$) when $q=1$. This limit identifies $k$ to Boltzmann
constant. It is straightforward to verify that $S$ has the same properties as $S_{GS}$.

For {\it microcanonical ensemble}, we have $S=\frac{k}{q}lnw$ which decreases with increasing $q$ value. In
general, $\Delta S=S-S_{GS}<0$ (or $>0$) if $q>1$ (or $q<0$) as shown in Fig.(1). This result is consistent
with the fact that the entropy of a lattice of chaotic maps increases with strong coupling\cite{Bati02}. The
reader will find later that decreasing $q$ implies increasing coupling for some correlated systems.

For {\it canonical ensemble}, the maximum entropy subject to the constraints from Eq.(\ref{3}) and (\ref{4})
leads to :

\begin{equation}
p_i=\frac{1}{Z}e^{-\beta E_i}                   \label{15}
\end{equation}
with $Z=\{ \sum_{i=1}^w e^{-q\beta E_i} \}^{1/q}$ where $E_i$ is the energy of the system at state $i$. The
Lagrange multiplier $\beta$ can be determined by $\frac{\partial S}{\partial U}=k\beta=\frac{1}{T}$ where $T$
is the absolute temperature and $U$ the internal energy of the system given by $U=\sum_{i=1}^{w}p_i^qE_i$. It
is easy to verify that $U=-\frac{\partial}{\partial\beta}\ln Z$.

For {\it grand canonical ensemble}, the average number of particles is calculated by
$\bar{N}=\sum_{i=1}^{w}p_i^qN_i$ which can be used as a constraint of maximum entropy. The partition function
is then given by $Z=\{ \sum_{i=1}^w e^{-q\beta (E_i-\mu N_i)} \}^{1/q}$ which gives
\begin{equation}                                \label{22b}
\bar{N}=\frac{1}{\beta}\frac{\partial}{\partial \mu} \ln Z
\end{equation}
For quantum particle systems, the usual procedures using Eq. (\ref{22b}), $E_i=\sum_k (n_k)_i\epsilon_k$ and
$N_i=\sum_k(n_k)_i$ will lead to
\begin{equation}                                \label{23}
\bar{n}_k=\frac{1}{e^{\omega\beta (\epsilon_k-\mu)}\pm 1}
\end{equation}
where $\bar{n}_k$ is the occupation number of the one-particle state $k$ of energy $\epsilon_k$ and $\mu$ the
chemical potential. "+" is for fermions and "-" for bosons.

\section{Examples of incompleteness effect}

Now I am presenting examples of the applications of EIS to some simple models. The reader should regard them as
demonstrations of the incompleteness behaviors of physical systems.

\subsection{``Ideal gas"}
The usual calculations give :

\begin{equation}                             \label{44}
Z=\{\frac{V}{h^3}[\frac{2\pi mkT}{q}]^{3N/2} \}^{1/q},
\end{equation}

\begin{equation}                            \label{45}
U=\frac{3}{2q}NkT
\end{equation}
and
\begin{equation}
(C_v)=\frac{3}{2q}Nk                                  \label{46}
\end{equation}
where $h$ is the Planck constant.

The effect of the incompleteness can be estimated through the energy difference
\begin{equation}                                    \label{47}
\Delta U=U-U_0=(\frac{1}{q}-1)\frac{3}{2}NkT
\end{equation}
which is positive (implying repulsion type interactions) for $q<1$ and negative (attraction type interactions)
for $q>1$ ($U_0$ is the internal energy of the conventional ideal gas when $q=1$).

The reader may be surprised by the treatment of ideal gas with a theory for systems including non negligible
interactions. From the viewpoint of IS, this is just a good example of the philosophy of generalized theory
with empirical parameter $q$ which is introduced to ``absorb" the effects of ``unaccessible" interactions or
their effects. In this way, interacting systems may be mathematically treated as noninteracting or conventional
ones including only the describable interactions. We are entitled to use, e.g., $p^2/2m$ for the total energy
of interacting ``free particle", where $p$ is the momentum and $m$ the mass of the particle. The extra
interaction energy is ``absorbed" in the parameter $q$ when it has different values from unity.

\subsection{Transport phenomena of ideal gas}

Let $W$ represent the number of particle (of mass $m$) collisions happening per second per unit volume, a usual
calculation \cite{Huang} will give :

\begin{equation}                                         \label{47a}
W=2n^2\sigma\sqrt{\frac{kT}{q\pi m}}= n^2\sigma \bar{v} \sqrt{\frac{2}{\pi}}
\end{equation}
where $n$ is the particle density and $\bar{v}=\sqrt{\frac{2kT}{qm}}$ is the most probable speed of a particle.
Let the mean free path of a particle be denoted by $\lambda$, the collision time $\tau$ (duration of $\lambda$)
is defined as follows :

\begin{equation}                                         \label{47b}
\tau=\frac{\lambda}{\bar{v}}=\frac{n}{2W}=\frac{1}{4n\sigma}\sqrt{\frac{q\pi m}{kT}}.
\end{equation}
In this framework, $\lambda$ does not change with respect to $BGS$ case, but $\tau$ is linked to $q$ and
increases with increasing $q$ value. This behavior of $\tau$ can affect the electrical conductivity $\sigma_e$
of metals with {\it free electron model} because

\begin{equation}                                         \label{47c}
\sigma_e= \frac{ne^2\tau}{m}=\frac{e^2}{4\sigma}\sqrt{\frac{q\pi}{mkT}}.
\end{equation}
which increases with increasing $q$ value.

\begin{figure}[h] \label{f2}
\includegraphics[width=5in,height=4in]{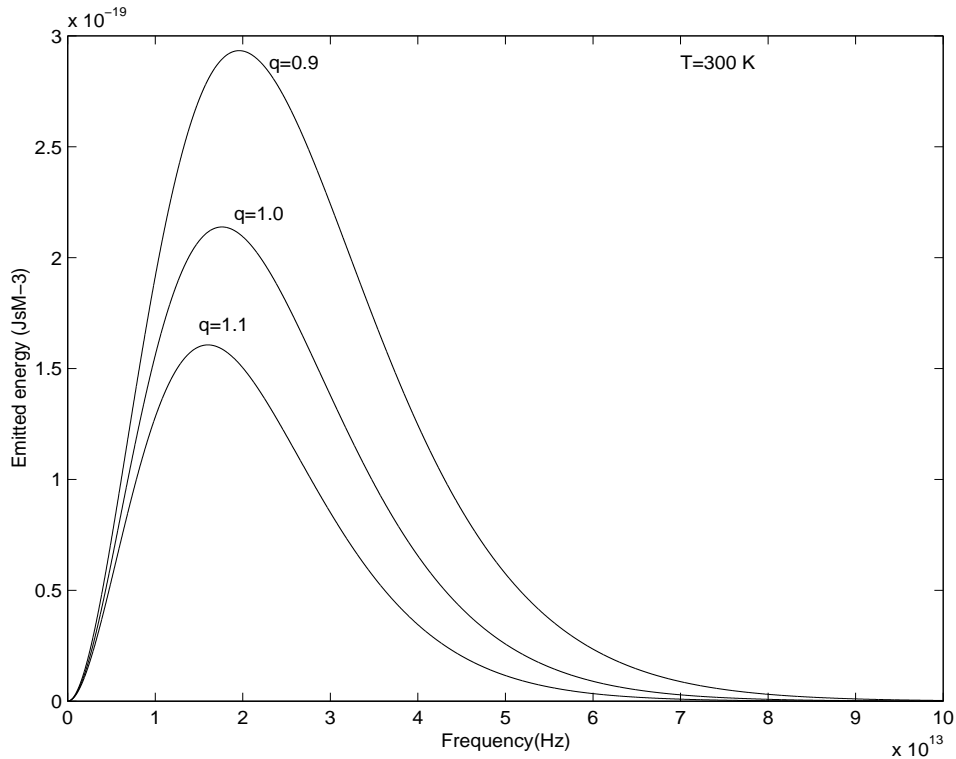}
\caption{Extensive generalized distribution of blackbody. It can be seen that the total emitted energy and the
maximal frequency of the emission increase when $q$ decreases.}
\end{figure}

\subsection{blackbody}

The generalized Planck law is given by

\begin{equation}                                   \label{48}
\rho_q(\nu,T)=\frac{8\pi h\nu^3}{c^3}\frac{1}{e^{q\beta h\nu}-1}
\end{equation}
where $\nu$ is the emission frequency and $c$ the light speed. The generalized Stephan-Boltzmann law is given
by $E_q(T)=\frac{\sigma}{q^4}T^4$ where $\sigma$ is the usual Stephan-Boltzmann constant. The distribution
Eq.(\ref{48}) is plotted in Figure 2 for different $q$ values to show the effect of the incompleteness of
information. We note that this EIS generalized blackbody is essentially different from the nonextensive
one\cite{Wang97}. Firstly, the high energy cutoff of nonextensive blackbody is absent in Eq.(\ref{48}).
Secondly, the Einstein emission and absorption coefficients can be easily shown to satisfy $B_{21}=B_{12}$ in
this work as in the case of the conventional Bose-Einstein theory. But with nonextensive blackbody, the ratio
$B_{21}/B_{12}$ varies from zero to infinity according to $q$ value\cite{Wang98}, which possibly opens a door
for the interpretation of the behavior of {\it micro chaotic laser}\cite{Chaoslaser} which still remains a
mystery.

\begin{figure}[p] \label{f3}
\includegraphics[width=5in,height=4in]{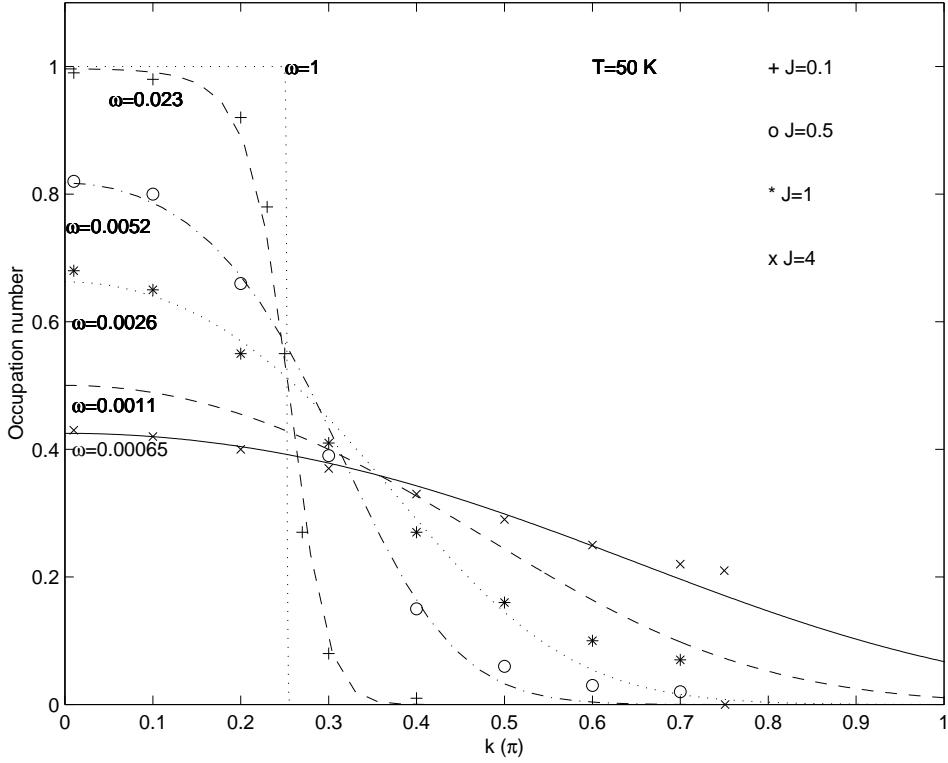}
\caption{Comparison of IS fermion distribution (lines) at various values of $q$ with the numerical results
(symbols) of Moukouri el al on the basis of Kondo lattice $t-J$ model for different coupling constant $J$[15].
We note the flattening of $n$-drop at Fermi energy $\epsilon_F$ with decreasing $q$ value. When $q\rightarrow
0$, we have $n=1/2$ for all possible states. When $q\rightarrow \infty$, we have $n=1$ for all states below
$\epsilon_F$ and $n=0$ for all states above $\epsilon_F$ for any temperature. In the calculations, the density
of electrons is chosen to give $k_{f_0}=0.35\pi$ in the first Brillouin zone. We see that the lines fit well
the numerical results for about $J<2$. When the coupling is stronger, a long tail in the KLM distributions
begins to develop at high energy and can not be reproduced by IS distribution. A new Fermi surface starts to
appear and a sharp $n$ drop (cutoff) takes place at about $k=0.7\pi$. These strong correlation effects are
absent in the present IS fermion distribution.}

\end{figure}

\subsection{Ideal fermion gas}
From the EIS fermion distribution given by Eq.(\ref{23}), it is easy to verify that the Fermi energy
$\epsilon_F^0$ at $T=0$ in the generalized version is the same as in the conventional Fermi-Dirac theory. The
zero temperature limit of the distribution is therefore not changed.

If $q$ is not very different from unity and $T$ is not too high, the derivative of $\overline{n}$ does not
vanish only when $\epsilon\approx\epsilon_F$. In this case, we can use Sommerfeld integral \cite{Sommer} to
calculate Fermi energy $\epsilon_F$ and internal energy for 3D fermion gas. The result is :

\begin{equation}                                        \label{55}
\epsilon_F\cong \epsilon_F^0[1-\frac{\pi^2}{12}(\frac{kT}{q\epsilon_F^0})^2],
\end{equation}
and
\begin{equation}                                        \label{56}
U\cong U_0[1+\frac{5\pi^2}{12}(\frac{kT}{q\epsilon_F^0})^2],
\end{equation}
where $U_0=\frac{3}{5}N\epsilon_F^0$ is the internal energy of fermion gas at $T=0$. The above equations show
that the decrease of $q$ leads to internal energy increase and a drop of Fermi level. This is because that,
when $q<1$, the particles are driven by the repulsion from the lower energy states to higher ones. We can see
from Eq.(\ref{23}) that, If $q=0$ with $T>0$, the repulsion would be so strong that all states are equally
occupied. On the contrary, if $q\rightarrow \infty$, all particles are constrained by the attraction to stay at
the lowest states, like the case of zero temperature distribution. This $q$ effect can be seen in Figure 3
where Eq.(\ref{23}) is plotted for 1-D fermion gas with different $q$ values smaller than unity.

The heat capacity $(C_v)_q$ and the magnetic susceptibility $\chi_q$ of {\it electron gas} can be given by :
\begin{equation}                                     \label{57}
(C_v)_q\cong U_q^0\frac{5\pi^2}{12}(\frac{k}{q\epsilon_F^0})^2T =\frac{\gamma_0}{q^2}T
\end{equation}
and
\begin{equation}
\chi_q\cong \chi_0[1-\frac{\pi^2}{12}(\frac{kT}{q\epsilon_F^0})^2].
                                                        \label{58}
\end{equation}
where $\gamma_0$ is the conventional coefficient of the electronic heat capacity and $\chi_0$ the conventional
susceptibility of electron gas at $T=0$.

With the help of Eq.(\ref{57}), the ratio of the effective mass $m_{th}$ to the mass $m$ of an electron can be
related to the parameter $q$ as follows :

\begin{equation}                                     \label{57a}
\frac{m_{th}}{m}=\frac{(C_v)_q(observed)} {C_v(theoretical)}=\frac{1}{q^2}.
\end{equation}
For heavy fermions, we surely have $q<1$. The reader will find below that this fact is confirmed by the $q$
values found for correlated electrons. We remember that above relationships are valid only for $q$ and $T$
values which keep the sharp $n$ drop at $\epsilon_F$.

On the other hand, in Figure 3, the calculations are valid for any $q$ value. For small $q$, we note a
flattening of the $n$-drop at $\epsilon_F$ with decreasing $q$ values. This is just a behavior of correlated
electrons observed in some experimental and numerical
results\cite{Corelec2,Corelec3,Corelec4,Ronn98,Puti98a,Puti98b}. Figure 3 shows a comparison of the momentum
distribution (lines) given by Eq.(\ref{23}) for 1-D fermions with numerical results (symbols) given by
numerical simulation based on Kondo lattice $t-J$ model (KLM)\cite{Corelec3}. We see that IS momentum
distribution fits well the numerical results for about $J<2$, the weak coupling regime\cite{Corelec3}. When the
coupling is stronger ($J>2$), a long tail in the KLM distributions begins to develop at high energy and can not
be fitted with present fermion distribution. In addition, in the strong coupling regime, a new Fermi surface
starts to appear and a sharp $n$ drop (cutoff) takes place at about $k=0.7\pi$. These strong correlation
effects are absent in the present EIS fermion distribution. Similar result is obtained with other numerical
results\cite{Corelec3,Wang02b}. This is consistent with the fact that high energy cutoff does not exist in EIS.
On the other hand, the sharp $n$ drop at a higher $\epsilon_F$ than the conventional one is indeed observed in
NIS fermion distribution\cite{Wang02a}. This result confirms our conjecture that EIS is only valid for weak
interaction cases due to its extensive nature and that nonextensive effect should be considered whenever
interactions become stronger.

\subsection{Conclusion}
From above results, we note that the introduction of parameter $q$ is equivalent to replacing $k$ in the
conventional $BGS$ theory by $k_q=k/q$. One may say that the incomplete $q$-normalization turns out to be a
renormalization of the constant without any change of physics. It should be noticed that the value of Boltzmann
constant has been determined for noninteracting systems. So any modification of this value undoubtedly implies
physical changes in the studied system as well as in the corresponding theory. The modification given by EIS is
a simple one as we guessed considering its validity for the cases where nonextensive effect of interactions can
be neglected. The successful fitting of the heavy fermion distributions shows that this modification reflects
at least a part of the reality and can be useful.

Summing up, the conventional BGS statistical mechanics is generalized for extensive systems on the basis of the
idea that we sometimes can not know all the informations about complex systems so that the physical or
observable probabilities become incomplete and do not sum to one. A generalized additive entropy is obtained by
using a so called {\it incomplete normalization} with a empirical parameter $q$ which is intended to ``absorb"
the effect of complex correlations and can be related to the energy of the studied system. So, in general,
$q<1$ and $q>1$ imply respectively repulsive and attractive effect of the complex correlations. This extensive
incomplete statistics is shown to be able to reproduce very well the quantum distributions of correlated heavy
electrons in weak coupling regime. On the other hand, EIS fails to describe strongly correlated conduction
electrons and localized $f$-electrons for which nonextensive effect should be taken into account.

\section{Acknowledgments}
I acknowledge with great pleasure the useful discussions with Professors A. Le M\'ehaut\'e, Dr. L. Nivanen and
Prof. J.P. Badiali on some points of this work.

\end{document}